\documentclass[a4paper]{revtex4}
\usepackage{graphicx}
\usepackage{fancyhdr}
\usepackage{amsmath}
\usepackage{color}

\begin{document}

\title{Probing Baryon Asymmetry of the Universe by Using Lepton
  Universality%
\footnote{This contribution is for a talk given by
T. Asaka at ``{\it Toyama International Workshop on Higgs as a Probe of New Physics 2015 (HPNP 2015)}''  (Toyama University, Japan, 
February 2015) and is based on the work~\cite{Asaka:2014kia}.}
}


\author{Takehiko Asaka}
\affiliation{Department of Physics, Niigata University, Niigata 950-2181, Japan}
\author{Shintaro Eijima}
\affiliation{Institut de Th\'eorie des Ph\'enom\`enes Physiques, \'Ecole Polytechnique
F\'ed\'erale de Lausanne, CH-1015 Lausanne, Switzerland}
\author{Kazuhiro Takeda}
\affiliation{Graduate School of Science and Technology, Niigata University, Niigata 950-2181, Japan}

\begin{abstract}
  We study the model with three right-handed neutrinos which masses
  are smaller than the weak scale ${\cal O}(10^2)$ GeV (called as the
  $\nu$MSM).  The model can explain the origin of neutrino masses by
  the seesaw mechanism, offer a candidate of dark matter and realize
  the baryogenesis via neutrino oscillation.  The seesaw mechanism at
  such energy scales can induce phenomenon which are observable by
  experiments.  As an example, we discuss the lepton universality of
  charged kaon decays in this model.  It is shown that the heavy
  neutral leptons accounting for the neutrino masses and the cosmic
  baryon asymmetry can give a significant correction to the lepton
  universality, and that the deviation from the Standard Model
  prediction can be large as ${\cal O}(10^{-3})$ which will be probed
  by near future experiments.
\end{abstract}

\maketitle

\thispagestyle{fancy}


%%%%%%%%%%%%%%%%%%%%%%%%%%%%%%%%%%
\section{Introduction}
Right-handed neutrinos are well-motivated particles beyond the Standard
Model (SM).  They can offer solutions to various problems inherent in
the SM.  When their Majorana masses are much heavier than the weak
scale $\Lambda_W ={\cal O}(10^2)$ GeV, they give a natural explanation
(the seesaw mechanism~\cite{Seesaw}) for tiny masses of active
neutrinos observed in oscillation experiments.  Remarkably, such
right-handed neutrinos can explain the origin of the baryon asymmetry
of the universe (BAU) by the leptogenesis~\cite{Fukugita:1986hr}.  The
canonical scenario of the leptogenesis requires that the Majorana masses
should be heavier than ${\cal O}(10^9)$ GeV.  It is then almost impossible
to test such particles by direct experiments in near future.

Here we consider another possibility, called as the $\nu$MSM (neutrino
Minimal SM)~\cite{Asaka:2005an,Asaka:2005pn}.  It is an extension of
the SM by three right-handed neutrinos 
$\nu_{R I}$ ($I=1,2,3)$ 
with Majorana masses $M_I$, which Lagrangian is given~by
\begin{eqnarray}
  {\cal L}
  = 
  {\cal L}_{\rm SM}
  +
  i \overline{\nu_{R I}} \gamma^\mu \partial_\mu \nu_{R I}
  - F_{\alpha I} \overline {L_\alpha} \Phi \nu_{R I}
  - \frac{M_I}{2} \overline{\nu_{R I}^c} \nu_{R I} + h.c. \,,
\end{eqnarray}
where ${\cal L}_{\rm SM}$ denotes the SM Lagrangian.
The model requires the hierarchies of neutrino masses
as $|M_D|_{\alpha I} = |F_{\alpha I}| \langle \Phi \rangle \ll M_I$ 
to realize the seesaw mechanism, and in addition 
Majorana masses are taken as  $M_I \lesssim \Lambda_W$.  
In this model three heavy neutral
leptons $N_I$ ($I=1,2,3)$ are present in addition to the usual three
active neutrinos $\nu_i$ ($i=1,2,3$).  The mixings of these states   
are represented as
$\nu_{L \alpha} = U_{\alpha i} \, \nu_i + \Theta_{\alpha I} \, N_I^c$
with the PMNS mixing matrix $U$ for active neutrinos
and $\Theta_{\alpha I} = F_{\alpha I} \langle \Phi \rangle /M_I$
for heavy neutral leptons.
The lightest one $N_1$ can be a
dark matter candidate, known as the so-called ``sterile neutrino dark
matter'' (see, for example, \cite{Boyarsky:2009ix}).  The heavier ones
$N_2$ and $N_3$ are responsible to the seesaw mechanism for neutrino
masses and also the baryogensis via neutrino
oscillation~\cite{Akhmedov:1998qx,Asaka:2005pn}.

One notable feature of the model is that heavy neutral leptons can be
tested by various experiments and also cosmological
observations~\cite{Kusenko:2004qc,Gorbunov:2007ak,Atre:2009rg}.  In
this talk based on the work~\cite{Asaka:2014kia} we discuss the
possible impacts of heavy neutral leptons to the two-body decays of
charged kaon and the testability of the $\nu$MSM by using the lepton
universality of kaon decays. (See Ref.~\cite{Asaka:2014kia} for the
lepton universality of charged pion decays.)

%%%%%%%%%%%%%%%%%%%%%%%%%%%%%%%%%%
\section{Lepton Universality in the $\nu$MSM}
We consider the lepton universality of charged kaon decays, which is
given by~\cite{Shrock:1980ct,Shrock:1981wq}
\begin{eqnarray}
  R_K = \frac{\Gamma (K^+ \to e^+ \nu)}{ \Gamma (K^+ \to \mu^+ \nu)} \,.
\end{eqnarray}
In the SM the expression of $R_K$ is estimated as
\begin{eqnarray}
  R_K^{\rm SM} = 
  \left( \frac{m_e}{m_\mu} \right)^2 
  \left( \frac{ m_K^2 - m_e^2} {m_K^2 - m_\mu^2} \right)^2 
  \left( 1 + \delta R_K \right) 
  \,,
\end{eqnarray}
where $\delta R_K$ denotes the radiative correction.  It should be
noted that the hadronic uncertainty in the width is canceled
considerably by taking the ratio, and then its theoretical prediction
is very precise $R_K^{\rm SM } = (2.477 \pm 0.001) \times
10^{-5}$~\cite{Finkemeier:1995gi,Cirigliano:2007xi}.  On the other
hand, the measurements of $R_K$ have been done at high precision by
experiments~\cite{Ambrosino:2009aa,Goudzovski:2010uk,NA62:2011aa,Lazzeroni:2012cx}.
The recent result gives $R_K^{\rm exp} = (2.488 \pm 0.010) \times
10^{-5}$~\cite{Lazzeroni:2012cx}.  It is found that the deviation of
the lepton universality from the SM prediction is small as
\begin{eqnarray}
  \Delta r_K = \frac{R_K}{R_K^{\rm SM}} -1 
  = (4 \pm 4) \times 10^{-3} \,,
\end{eqnarray}
and hence it can be used as the significant probe for
physics beyond the SM.  (See, for example, the recent analysis
in Refs.~\cite{Abada:2012mc,Abada:2013aba}.)

In the $\nu$MSM, $K^+$ decays into not only active neutrinos
but also heavy neutral leptons if kinematically allowed
\begin{eqnarray}
  R_K = 
  \frac{
    \sum_{i=1,2,3} \Gamma (K^+ \to e^+ \nu_i)
    + 
    \sum_{I=1,2,3} \Gamma (K^+ \to e^+ N_I)
  }{
    \sum_{i=1,2,3} \Gamma (K^+ \to \mu^+ \nu_i)
    + 
    \sum_{I=1,2,3} \Gamma (K^+ \to \mu^+ N_I)
  }  \,,
\end{eqnarray}
and the deviation is  given by~\cite{Shrock:1980ct}
\begin{eqnarray}
  \Delta r_K
  =
  \frac{
    \sum_{i=1,2,3} |U_{e i}|^2 +
    \sum_{I=1,2,3} |\Theta_{e I}|^2 G_{e I}
  }{
    \sum_{i=1,2,3} |U_{\mu i}|^2 +
    \sum_{I=1,2,3} |\Theta_{\mu I}|^2 G_{\mu I}
  }
  - 1 \,,
  \label{eq:DelrK0}
\end{eqnarray}
where $G_{\alpha I} = 0$ if $M_I > m_K - m_{\ell_\alpha}$;
and $G_{\alpha I} =
  \frac{ r_\alpha + r_I - (r_\alpha - r_I)^2 }
  { r_\alpha ( 1 - r_\alpha)^2}
  \sqrt{ 1 - 2 (r_\alpha + r_I) +(r_\alpha - r_I)^2 } $
otherwise.
%\begin{eqnarray}
%  G_{\alpha I} =
%  \frac{ r_\alpha + r_I - (r_\alpha - r_I)^2 }
%  { r_\alpha ( 1 - r_\alpha)^2}
%  \sqrt{ 1 - 2 (r_\alpha + r_I) +(r_\alpha - r_I)^2 } \,,
%\end{eqnarray}
Here $r_\alpha = m_{\ell_\alpha}^2/m_K^2$ and $r_I = M_I^2 /m_K^2$. 
We should note that
the mixing elements of active neutrinos and heavy neutral leptons
satisfy the unitarity condition
\begin{eqnarray}
  \label{eq:Unitarity}
  \sum_{i=1,2,3} |U_{\alpha i}|^2
  + \sum_{I=1,2,3} |\Theta_{\alpha I}|^2 = 1 \,,
\end{eqnarray}
and hence we can write $\Delta r_K$  as
\begin{eqnarray}
  \Delta r_K
  =
  \frac{
    1 + 
    \sum_{I=1,2,3} |\Theta_{e I}|^2 
    \left[ G_{e I} - 1 \right]
  }{
    1 +
    \sum_{I=1,2,3} |\Theta_{\mu I}|^2 
    \left[ G_{\mu I} - 1 \right]
  }
  - 1
  \,.
  \label{eq:delRk}
\end{eqnarray}
Therefore, we find that the deviation $\Delta r_K$ in the $\nu$MSM is
determined by the masses $M_I$ and mixing elements $\Theta_{\alpha I}$
of heavy neutral leptons.  It has been shown in Ref.~\cite{Asaka:2014kia} that
heavy neutral lepton $N_1$ gives a negligible contribution to $\Delta
r_K$ (\ref{eq:delRk}) by imposing the severe constraints on
$\Theta_{\alpha 1}$ from cosmological observations and we shall
neglect it from now on.  Further, $N_2$ and $N_3$ should be
quasi-degenerate in mass to generate enough BAU, and 
we take $M_2 = M_3 = M_N$ since the mass difference
has no significant effect on $\Delta r_K$~\cite{Asaka:2014kia}.

Let us first consider the case when $M_N < m_K - m_\mu$ (both $K^+ \to
\mu^+ N_I$ and $K^+ \to e^+ N_I$ are kinematically allowed).  It is
interesting to note that the decay width of $K^+ \to \ell_\alpha^+N_I$
is suppressed by $|\Theta_{\alpha I}|^2$, but is enhanced by
$(M_N/m_{\ell_\alpha})^2$ compared with $K^+ \to \ell_\alpha^+
\nu_\alpha$ due to the helicity suppression~\cite{Shrock:1980ct}.
Since this enhancement factor is much larger for the decay into $e^+$
than that into $\mu^+$, the considering model
gives a positive $\Delta r_K$
\begin{eqnarray}
  \label{eq:DELRK_ap}
  \Delta r_K \simeq \sum_{I=2,3} | \Theta_{eI} |^2 \,
   \frac{M_N^2}{m_e^2}
  \left(
    1- \frac{M_N^2}{m_K^2} \right)^2 \,.
\end{eqnarray}
It is seen that $\Delta r_K$ is bounded from above by taking into
account the experimental upper bounds on the mixing elements
$|\Theta_{e I}|$.  PS191 experiment gives the severest bound $|\Theta_{e I}|^2 < {\cal
  O}(10^{-9})$--${\cal O}(10^{-8})$ for $M_N \simeq 200$--400~MeV~\cite{Bernardi:1985ny}, and then it is expected that
$\Delta r_K <{\cal O}(10^{-4})$--${\cal
  O}(10^{-3})$ by considering the enhancement factor of
$(M_N/m_e)^2 \sim10^{5}$.
When $K^+ \to \mu^+ N_I$ is forbidden,
the correction $\Delta r_K$ is very similar to the above case
as long as $K^+ \to e^+ N_I$ is open.

On the other hand, even when $K^+ \to e^+ N_I$ is kinematically forbidden,
the deviation $\Delta r_K$ can be present because of the non-unitarity 
of the PMNS matrix (see Eq.~(\ref{eq:Unitarity})) as
\begin{eqnarray}
  \label{eq:DEL_rK_h}
  \Delta r_K \simeq \sum_{I = 2,3} 
  \bigl( |\Theta_{\mu I}|^2 - |\Theta_{e I}|^2 \bigr) \,.
\end{eqnarray}
In this case the sign of $\Delta r_K$ is determined according to the
relative sizes of $|\Theta_{\mu I}|^2$ and $|\Theta_{e I}|^2$ and the
magnitude is $|\Delta r_K |\lesssim |\Theta|^2 = {\cal O}(10^{-9})$--${\cal
  O}(10^{-7})$.

\begin{figure}[t]
  \centerline{
  \includegraphics[width=8cm]{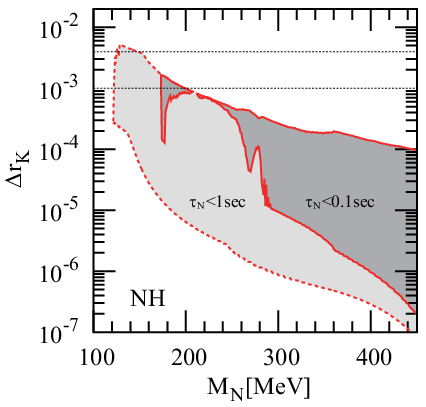}%
  \includegraphics[width=8cm]{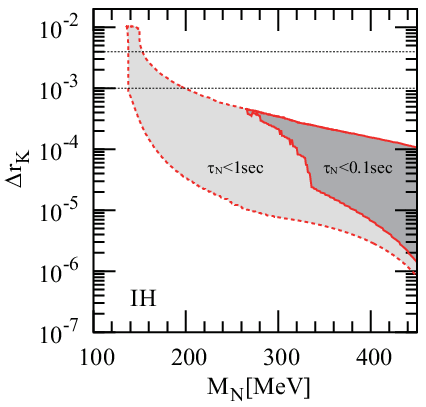}%
  }%
  \caption{\it
    $\Delta r_K$ in the $\nu$MSM for the NH case (left panel) 
    and IH case (right panel)~\cite{Asaka:2014kia}.
    Possible regions are shown by the shaded regions with 
    red-solid line or red-dashed line for the case with
    the cosmological lifetime bound $\tau_{N_{2,3}}< 0.1$ s
    or $\tau_{N_{2,3}}< 1$ s.  
    The horizontal (black dotted) lines are 
    $\Delta r_K = 4 \times 10^{-3}$ (current central value~\cite{Lazzeroni:2012cx})and 
    $\Delta r_K = 10^{-3}$ (which will be reached by
    the near future experiments).
  }
  \label{fig:DEL_RK_1sec}
\end{figure}

Now, we are at the point to present the numerical prediction of
$\Delta r_K$ in the $\nu$MSM by using the exact formula
(\ref{eq:DelrK0}).  In this analysis we impose the constraints from
direct search experiments and cosmological lifetime bound
$\tau_{N_{2,3}} < 0.1$ s~\cite{Dolgov:2000pj}.  (See the details in
Ref.~\cite{Asaka:2014kia}.)  The predicted range of $\Delta r_K$ by
varying all the free parameters is shown in
Fig.~\ref{fig:DEL_RK_1sec}. 
We should mention that, as demonstrated in Refs.~\cite{Canetti:2010aw,Asaka:2013jfa},
enough BAU can be generated by $N_2$ and $N_3$ in 
this mass region.
 It is found that $\Delta r_K ={\cal
  O}(10^{-7})$--${\cal O}(10^{-3})$ for the NH case, and $\Delta r_K =
{\cal O}(10^{-6})$--${\cal O}(10^{-3})$ for the IH case, where we have
considered $M_N < 450$ MeV and $\tau_{N_{2,3}}<0.1$ s.  The predicted
region becomes wider if the lifetime bound is relaxed as
$\tau_{N_{2,3}}<1$ s.
%%%%%%%%%%%%%%%%%%%%%%%%%%%%%%%%%%%%%%%%%%%%%%%%%%%%%%%%%%%%%%%%%%%%%%%%%  
%%%%% ** Figure ** %%%%%%%%%%%%%%%%%%%%%%%%%%%%%%%%%%%%%%%%%%%%%%%%%%%%%% 
%%%%%%%%%%%%%%%%%%%%%%%%%%%%%%%%%%%%%%%%%%%%%%%%%%%%%%%%%%%%%%%%%%%%%%%%%

%%%%%%%%%%%%%%%%%%%%%%%%%%%%%%%%%%
\section{Conclusions}
We have investigated the lepton universality of charged kaon
decays in the framework of the $\nu$MSM.  Among three heavy neutral leptons of the
model, $N_2$ and $N_3$, which are responsible to the seesaw mechanism
of neutrino masses and the baryogenesis via neutrino oscillation, can
give significant contributions to kaon decays and modify the ratio
$R_K$ of the partial widths $K^+ \to e^+ \nu$ and $K^+ \to \mu^+ \nu$
from the prediction by the SM.

It has been shown that the deviation of the lepton universality 
in kaon decays can be large as $\Delta r_K = {\cal O}(10^{-3})$
even if we apply the constraints on the mixing elements
for heavy neutral leptons $N_2$ and $N_3$ from the direct search experiments
and the lifetime bound $\tau_{N_{2,3}} < 0.1$ s from the Big Bang nucleosynthesis.
Such a large deviation can be obtained when the mass is
$M_N \sim 180$ MeV. (Such a large value can also be 
obtained when $M_N$ is just above 450 MeV~\cite{Asaka:2014kia}.)
We have found that the sign of $\Delta r_K$ is positive
if the decays $K^+ \to e^+ N_{2,3}$ are open.
Note that, if we relax the lifetime bound as $\tau_{N_{2,3}} < 1$ s,
then the deviation can be large as ${\cal O}(10^{-2})$.

Interestingly, such a large deviation of the lepton universality
in kaon decays will be explored by near future experiments
as NA62~\cite{Goudzovski:2012gh} and TREK/E36~\cite{Kohl:2013rma} experiments.  Thus, these facilities
will offer the test for the origin of neutrino masses 
and BAU via physics of heavy neutral leptons $N_2$ and $N_3$ in the $\nu$MSM.

%%%%%%%%%%%%%%%%%%%%%%%%%%%%%%%%%%
\begin{acknowledgments}
T.A. was supported by JSPS KAKENHI Grant Number 25400249 and 26105508, 
and S.E. was supported by Swiss National Science Foundation.
T.A. expresses thanks to the organizers of HPNP 2015 for their warm 
hospitality.
\end{acknowledgments}

\bigskip % extra skip inserted

\end{document}